\documentclass{an}
\usepackage{graphicx}
\usepackage{times}
\Volume{00}              
\Year{0000}              
\Month{00}               
\Pagespan{000}{000}      

\begin{document}
\lhead[\thepage]{M. Verheijen: The Disk Mass project}
\rhead[Astron. Nachr./AN~{\bf XXX} (200X) X]{\thepage}
\headnote{Astron. Nachr./AN {\bf 32X} (200X) X, XXX--XXX}

\title{The Disk Mass project; science case for a new PMAS IFU module}

\author{M.A.W.~Verheijen$^1$ \and M.A.~Bershady$^2$ \and D.R.~Andersen$^3$ \and
  R.A.~Swaters$^4$ \and K.~Westfall$^2$ \and A.~Kelz$^1$ \and M.M.~Roth$^1$}
\institute{
Astrophysikalisches Institut Potsdam, An der Sternwarte 16, 14482 Potsdam, Germany \and
University of Wisconsin, Dept. of Astronomy, 475N Charter Street, Madison, WI 53706, U.S.A. \and
Max-Planck-Institut f\"ur Astronomie,  K\"onigstuhl 17, 69117 Heidelberg, Germany \and 
Johns Hopkins University, Dept. of Physics and Astronomy, 3400 North Charles
Street, Baltimore, MD 21218, U.S.A.}

\date{Received {date will be inserted by the editor}; 
accepted {date will be inserted by the editor}} 

\abstract{We present our Disk Mass project as the main science case for
  building a new fiber IFU-module for the PMAS spectrograph, currently mounted
  at the Cassegrain focus of the 3.5m telescope on Calar Alto. Compared to
  traditional long-slit observations, the large light collecting power of
  2-dimensional Integral Field Units dramatically improves the prospects for
  performing spectroscopy on extended low surface brightness objects with high
  spectral resolution. This enables us to measure stellar velocity dispersions
  in the outer disk of normal spiral galaxies. We describe some results from a
  PMAS pilot study using the existing lenslet array, and provide a basic
  description of the new fiber IFU-module for PMAS.
}
\correspondence{mv@aip.de}

\maketitle

\section{Introduction}

A major roadblock in testing galaxy formation models is the disk-halo
degeneracy; density profiles of dark matter haloes as inferred from rotation
curve decompositions depend critically on the adopted M/L of the disk
component (Figure 1). An often used refuge to circumvent this degeneracy is
the adoption of the maximum-disk hypothesis (van Albada \& Sancisi 1986).
However, this hypothesis remains unproven. Recently, Bell \& de Jong (2001)
have shown that stellar population synthesis models yield plausible {\it
  relative} measurements of stellar M/L in old disks, but uncertainties in the
IMF prevent an {\it absolute} measurement of total disk M/L from photometry.
Another tool to determine the M/L, and specifically whether disks are maximal,
is the Tully-Fisher relation, e.g. by looking for offsets between barred
vs. un-barred galaxies, but this too is only a {\it relative} measurement.
Evidently, none of these methods are suited to break the degeneracy, and
without an independent measurement of the M/L of the stellar disk, it is not
possible to derive the structural properties of dark matter haloes from
rotation curve decompositions.

An absolute measurement of the M/L can be derived from the vertical
component $\sigma_{\rm z}$ of the stellar velocity dispersion. For a locally
isothermal disk, $\sigma_{\rm z} = \sqrt{\pi G(M/L)\mu z_o}$, with $\mu$ the
surface brightness, and $z_0$ the disk scale height. The latter is
statistically well-determined from studies of edge-on galaxies (de Grijs \&
van der Kruit 1996; Kregel et al. 2002). Thus, $\sigma_{\rm z}$ provides a
direct, kinematic estimate of the M/L of a galaxy disk and can break the
disk-halo degeneracy.

This kinematic approach has been attempted before (e.g. Bottema 1997), but
with long-slit spectroscopy on significantly inclined galaxies. These
observations barely reached out to 1.5 disk scale-lengths, required
significant radial binning, and because of the high inclinations of these
galaxies, the measured velocity dispersions required large and uncertain
corrections for the tangential ($\sigma_\phi$) and radial ($\sigma_{\rm r}$)
components of the stellar velocity dispersion ellipsoid.

With the advent of 2-dimensional spectroscopy using Integral Field Units
(IFU), the observational prospects for measuring $\sigma_{\rm z}$ have
improved dramatically. This is because a wide-field
(1$^\prime$$\times$1$^\prime$) IFU can collect and pipe much more light to a
spectrograph than a single 2$^{\prime\prime}$ wide long-slit.  The power of
IFUs with face-on galaxies lies in the ability to azimuthally average many
fibers. This yields clean $\sigma_{\rm z}$ measurements well beyond 2 disk
scale-lengths where contamination from bulge stars is negligible and the
rotation curve of the stellar disk has reached its flat part.
Here, we describe our Disk Mass project which employs two custom built IFUs to
measure $\sigma_{\rm z}$.

\section{The Disk Mass project}

\begin{figure}
\resizebox{\hsize}{!}
{\includegraphics[]{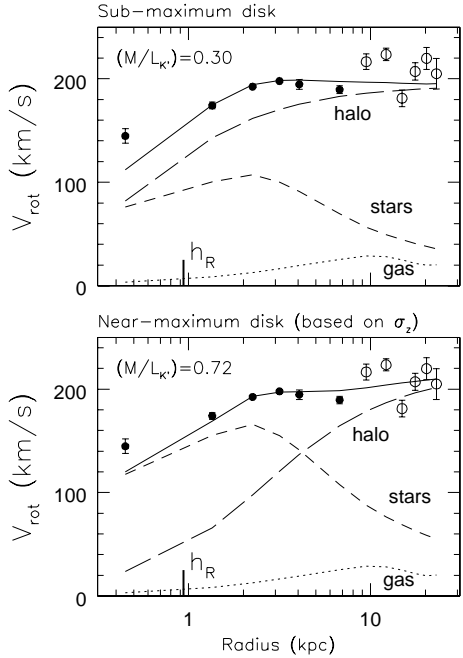}}
\caption{Different decompositions of the rotation curve of NGC 3982.
  Upper panel: assuming an arbitrary value for the stellar (M/L). Lower panel:
  using a stellar (M/L) based on re-evaluated $\sigma_{\rm z}$ measurements with
  SparsePak using the CaII lines. Both cases imply different density profiles
  for the dark matter halo (long-dashed curves).}
\label{figlabel}
\end{figure}

Measuring the vertical stellar velocity dispersion in the outer parts of
galaxy disks requires spectroscopy with a spectral FWHM resolution of
R$\approx$10$^4$ on objects with a surface brightness as low as $\mu$(B)=24.5
mag/arcsec$^2$, the typical value at three disk scale-lengths from the center
of a normal spiral galaxy.  With conventional long-slit spectrographs, this
can only be achieved with exceedingly long integration times of several nights
per object. However, it is our ambition to measure $\sigma_{\rm z}$ out to
three disk scale-lengths in a statistically significant sample of $\sim$40
kinematically well-behaved spiral galaxies, selected from a larger parent
sample of $\sim$100 nearly face-on galaxies with high quality H$\alpha$
velocity fields.

Secondary science goals of our survey include using the kinematic data for
constraining the shape of galactic potentials. Our stellar and gaseous
velocity fields will reveal kinematic perturbations due to non-circular halo
shapes and disk asymmetries (lopsidedness, spiral arms and disk ellipticity).
Previous studies have shown that such asymmetries may explain most of the
scatter in the Tully-Fisher (TF) relation (e.g. Rix \& Zaritsky 1995; Andersen
et al.  2001). Due to the favorable projection of the photometric disk
structure, our parent sample is uniquely suited for this study. Furthermore,
with our $\sigma_{\rm z}$ measurements we will also be able to calibrate the
mass scale of stellar population models.

Our ambition to measure $\sigma_{\rm z}$ in $\sim$40 galaxies required the
development of some special purpose instrumentation as well as a strategic
long-term observing program. We have taken the three following preparatory
steps:

First of all, to obtain a clean measurement of $\sigma_{\rm z}$, one needs to
minimize contributions from the radial and tangential components of the
velocity dispersion ellipsoid. For this reason we choose to observe nearly
face-on galaxies in which $\sigma_{\rm r}$ and $\sigma_\phi$ are almost
perpendicular to the line of sight and the observed velocity dispersion is
largely dominated by $\sigma_{\rm z}$. Since we are also interested in a
galaxy's total mass and the shape of its rotation curve, galaxies should not
be too face-on in order to derive kinematic inclinations based on H$\alpha$
velocity fields. Andersen and Bershady (2004) have proven to be able to derive
kinematic inclinations from regular, high signal-to-noise H$\alpha$ velocity
fields of galaxies as little inclined as 15 degrees; they constructed a
face-on TF relation based on kinematic inclinations, and found a
scatter similar to TF studies based on samples of more inclined galaxies.

Second, to measure $\sigma_{\rm z}$, a high spectral resolution is required,
given the expected low velocity dispersions in the outer disks of spiral
galaxies. Existing IFU spectrographs either lack the required light-gathering
power, spectral resolution, or spectral coverage. For example, SAURON's
instrumental {\it dispersion} is 105 km/s (Bacon et al.  2001); too large for
measuring $\sigma_{\rm z}$. INTEGRAL's spectral resolution of R$\approx$4200
or $\sim$70 km/s when using the largest fibers and the Echelle is also just
inadequate. The IFU instrumentation at the VLT aims at high spatial
resolutions. Consequently, individual spatial elements have become too small
to collect sufficient light for R$\approx$10$^4$ spectroscopy of low surface
brightness objects, despite the 8.4m diameter mirrors.

To obtain measurements of $\sigma_{\rm z}$ with sufficient spectral
resolution, Bershady et al. (2004a, 2004b) built the SparsePak IFU for the
WIYN facility at Kitt Peak. It consists of 75 bare object fibers in a sparsely
filled but regular hexagonal grid with a
72$^{\prime\prime}$$\times$71$^{\prime\prime}$ field of view. This grid can be
filled with three offset pointings. Each fiber has an extremely large diameter
of 4.69$^{\prime\prime}$ and collects light from a circular area of 17.3
arcsec$^2$ on the sky. Using the Echelle grating on the WIYN Bench
Spectrograph, a spectral resolution of R=1.1$\times$10$^4$ can be achieved
($\sim$40 km/s in the CaII region around $\sim$860nm).

Although the SparsePak IFU has great light collecting power, the Bench
Spectrograph to which the fibers pipe their collected light has a very low
throughput of $\sim$4\%. Consequently, we can use SparsePak only for
efficiently observing H$\alpha$ velocity fields, while stellar velocity
dispersion measurements are limited to the bright inner regions of spiral
galaxies.

\begin{figure}
\resizebox{\hsize}{!}
{\includegraphics[]{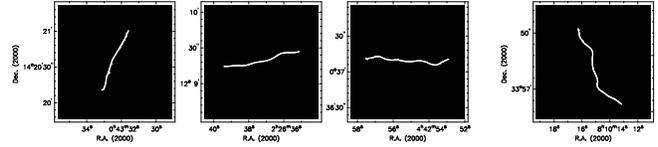}}
\caption{Several examples of H$\alpha$ velocity fields obtained with the
  SparsePak IFU on the WIYN telescope. The three velocity fields on the left
  are sufficiently regular to yield kinematic inclinations. The velocity field
  on the right is too irregular and UGC 4256 is not suitable for follow-up
  measurements of $\sigma_{\rm z}$ with PMAS.}
\label{figlabel}
\end{figure}

Third, the high throughput of the modularly built PMAS spectrograph (Roth et
al. 2000) makes it particularly attractive to equip it with a SparsePak-like
IFU module. A succesful pilot study with PMAS in March 2003 demonstrates that
PMAS provides the required throughput, spectral resolution and stability.
Equipped with the new fiber IFU module described in Section 4, it will be
possible with PMAS to measure $\sigma_{\rm z}$ in the dim outer regions of
disk galaxies.

\begin{figure}
\resizebox{\hsize}{!}
{\includegraphics[]{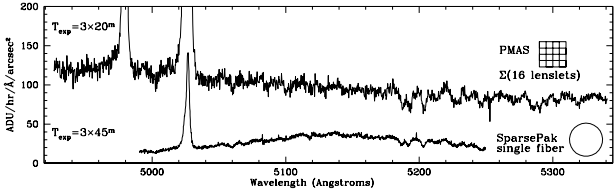}}
\caption{Comparison of the throughputs of PMAS and SparsePak based on
  observations of the same area in NGC 3982. Note the larger wavelength
  coverage of PMAS and the spectral vignetting of the SparsePak
  spectrum. The PMAS spectrum is the sum of 16 spectra from 16 lenslets
  to cover the same area as a single SparsePak fiber.}
\label{figlabel}
\end{figure}

Having built the new SparsePak and PMAS IFUs, we are now ready to measure
$\sigma_{\rm z}$ with an optimized experimental design, taking a two-phased
observational approach toward measuring galaxy disk masses over a 3-4 year
period.

\noindent
(i) We selected our parent sample from the UGC, choosing disk galaxies at
$|b|\ge 25^\circ$, with diameters of 1$^\prime$$-$1.5$^\prime$ to match
them to the field-of-views of SparsePak and PMAS, and with approximate
inclinations in the range 15$<$i$<$35, based on their optical axis ratios.
This yielded a total sample of 470 galaxies from which we remove the strongly
barred and interacting galaxies and randomly pick $\sim$100 galaxies to
observe.

\noindent
(ii) Using SparsePak, we are collecting high signal-to-noise H$\alpha$
velocity fields for $\sim$100 galaxies (Figure 2).  We need such a large
sample not only for a thorough statistical analysis of disk asymmetries and
their relation to the scatter in the TF relation, but also because we found
during our SparsePak pilot study that only 2 in 5 galaxies have sufficiently
regular H$\alpha$ velocity fields to determine both the rotation curve and
kinematic inclination. Because of intrinsic ellipticities of stellar disks
(Andersen et al. 2001), optical axis ratios do not provide good estimates of
the inclination, and a kinematic inclination is needed. Collecting these
H$\alpha$ data requires significant amounts of bright time granted by the
University of Wisconsin and matched by NOAO via an approved long-term program.
At the moment of this writing, some 70 H$\alpha$ velocity fields have already
been acquired.  They are inspected for their regularity, and kinematic
inclinations are being measured.

\noindent
(iii) From this large parent sample we will select $\sim$40 kinematically
regular galaxies suitable for follow-up measurements of $\sigma_{\rm z}$ with
PMAS. This number is needed for a statistically meaningful sampling of a range
in galaxy luminosity, colour, and disk surface brightness. With PMAS we will
focus on the MgI$b$ region of the spectrum ($\sim$515nm) where only few sky
lines are present. Comparing the H$\alpha$ and stellar kinematics yields a
measurement of the asymmetric drift which provides another means to constrain
$\sigma_{\rm z}$. Notably, the [OIII] emission line is also measured with
PMAS, yielding gas and stellar kinematics from the same observations.

\noindent
(iv) Galaxies with observed H$\alpha$ velocity fields are being imaged in the
U, B, V, R and I passbands on the 2.1m telescope on Kitt Peak. At the moment,
38 galaxies have been imaged through all five filters and 17 additional
galaxies have been observed in V and R. These photometric images will be used
to verify the pointings of our IFU observations, to study optical asymmetries,
to construct TF relations in different bands, and to calibrate stellar
population models.

SparsePak provides a high spectral resolution in the CaII region, and for
selected targets we may explore systematic differences between values of
$\sigma_{\rm z}$ measured with PMAS in the MgI$b$ region and with SparsePak in
the CaII region of the spectrum. This allows us to test the effects of radial
stellar population gradients, and population differences among galaxies.
Detecting the presence of a kinematically distinct thick disk requires an
additional increase in the signal-to-noise to be able to characterize the
detailed {\it shape} of the line-of-sight velocity distribution function,
beyond measuring its dispersion $\sigma_{\rm z}$. Currently, this
goal seems unattainable.

\section{PMAS test observations}

\begin{figure}[t]
\resizebox{\hsize}{!}
{\includegraphics[]{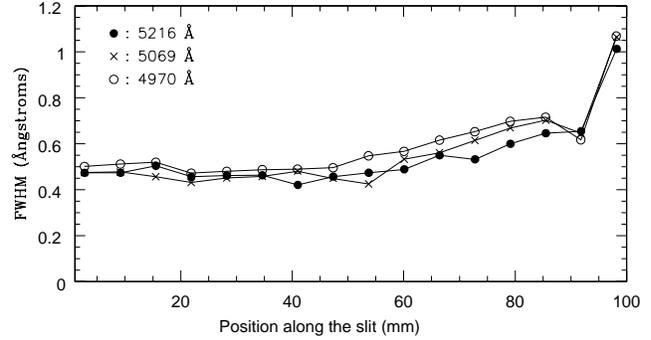}}
\caption{FWHM spectral resolution as a function of wavelength and position
  along the slit based on the existing 100$\mu$m PMAS fibers. In this case, the
  focus was only optimized for the center of the CCD.}
\label{figlabel}
\end{figure}

The PMAS spectrograph, with its high throughput optimized for blue and visual
wavelengths, provides us with the primary survey engine to measure
$\sigma_{\rm z}$ in the MgI$b$ region of the spectrum.  It is mounted on the
3.5m telescope at Calar Alto which has the same mirror diameter as the WIYN
telescope. Furthermore, PMAS has a modular design, and without too much
effort, the existing lenslet array can be exchanged with a new wide-field
SparsePak-like IFU module. To verify the usability of PMAS for measuring
$\sigma_{\rm z}$, we performed pilot observations on 2003 March 4-6 using the
existing lenslet array, configured for a 16$\times$16 arcsec field-of-view or
a light collecting area of 1 arcsec$^2$ per lenslet.  We were interested in
the achievable spectral resolution, throughput and stability of PMAS.

To verify the throughput, we observed the same galaxy with PMAS for which data
were collected earlier with SparsePak. One SparsePak fiber has the same
collecting area as $\sim$16 PMAS lenslets and we added the signal of 16 PMAS
spectra from lenslets located near the same position in the galaxy as one of
the SparsePak fibers. The detected signal of both the SparsePak and the
co-added PMAS spectra were scaled to the same exposure time, dispersion and
collecting area.  As illustrated in Figure 3, the SparsePak spectrum suffers
from significant spectral vignetting while the PMAS spectrum covers a larger
wavelength range, including more stellar absorption features. The throughput
of PMAS is a factor 2-8 higher than that of the WIYN Bench Spectrograph,
demonstrating that PMAS is roughly a factor 5 more efficient.

To achieve the required spectral resolution at 515nm, we used the I1200
grating, blazed for $\lambda$=1$\mu$m, in second order. By mounting the
grating backwards in its cartridge, we could rotate it to such a large angle
that an anamorphic demagnification of a factor 0.49 was obtained.  For the
existing 100$\mu$m fibers, we measured a FWHM spectral resolution of
$\sim$0.45\AA\ near the center of the CCD (Figure 4). This corresponds to
R$\approx$11400 at $\lambda$=5150\AA\ which implies that we can use 150$\mu$m
fibers for the new IFU module and still achieve an acceptable FWHM spectral
resolution of R$\approx$7600 or 39 km/s.

\begin{table}
\caption{Results of the flexure tests; relative shifts in pixels of a bright
  emission line in the spectral and spatial directions on the chip.}
\label{tablabel}
\begin{tabular}{r|rrrrrrr}\hline

 Dec. &  \multicolumn{7}{c}{Hour Angle}  \\ 
(deg) & $-6$ & $-4$ & $-2$ & $0$ & $+2$ & $+4$ & $+6$ \\

\hline
\multicolumn{8}{l}{Spectral shifts:} \\
 $+80$ & -2.42 & -1.86 & -1.97 &     1.23 & 2.95$^\dagger$ & 1.75 &  2.48 \\
 $+60$ &       & -2.57 &       &     0.82 &       & 1.40 &       \\
 $+40$ &       & -2.27 & -2.25 &     0.77 & 1.87$^\ddagger$ & 0.71 &       \\
 $+20$ &       & -2.72 &       &     0.34 &       & 0.21 &       \\
    0  &       & -2.63 & -2.46 &{\bf 0.00}& -0.12 & 0.21 &       \\
 $-20$ &       &       &       &    -0.30 &       &      &       \\

\hline
\multicolumn{8}{l}{Spatial shifts:} \\
 $+80$ &  4.79 &  4.59 &  4.39 &  1.91 & -0.64$^\dagger$ & -1.17 & -3.49 \\
 $+60$ &       &  4.64 &       &  0.95 &        & -4.00 &       \\
 $+40$ &       &  4.62 &  4.20 & -0.70 & -3.34$^\ddagger$ & -4.01 &       \\
 $+20$ &       &  4.66 &       & -0.77 &        & -3.71 &       \\
    0  &       &  4.65 &  4.16 &{\bf 0.00}& -2.41  & -3.54 &       \\
 $-20$ &       &       &       &  0.21 &        &       &       \\

\hline
\multicolumn{8}{l}{$^\dagger$out of focus, $^\ddagger$ after refocus} \\

\end{tabular}
\end{table}

Since PMAS is not mounted on an optical bench but located at the Cassegrain
focus, possible flexure is a main concern. To characterize the flexure we took
several exposures with the Thorium-Argon lamp, pointing the telescope at
various declinations and hour angles. For each pointing direction, we measured
the position of a bright emission line on the CCD. The shift in its centroid
with respect to a pointing at $\delta$=0$^\circ$, HA=0$^{\rm hr}$ is listed in
Table~1. We see that the flexure is quite constant on either side of
the meridian. Degradation of the spectra due to flexure is minimal during a 1
hour exposure, as long as no meridian transit occurs while exposing.
This can be avoided with a careful selection of our targets.

From the test observations we learned that PMAS provides the required
throughput, spectral resolution and flexure stability, and thus has the
capabilities to become an efficient instrument for measuring stellar velocity
dispersions. On the basis of these test results it was decided to build a new
SparsePak-like fiber IFU module for the PMAS spectrograph.

\section{PPak: a new fiber IFU-module for PMAS}

PPak consists of 331 bare fibers in a hexagonal grid behind a f/3.5 focal
reducer with a plate scale of 17.89 $^{\prime\prime}$/mm. The hexagonal
field-of-view has a diameter of 75$^{\prime\prime}$ while each fiber
is 2.68$^{\prime\prime}$ in diameter. An additional 36 fibers are placed
62$^{\prime\prime}$ from the center (Figure 5). Fifteen extra fibers are
diverted from the focal plane to be illuminated by
calibration lamps, keeping track of flexure and allowing for a synchronous
spectral calibration. Compared to a SparsePak fiber, each PPak fiber
collects only a third of the light, but there are 4.4$\times$ more fibers
feeding a highly efficient spectrograph.

Construction of PPak at the AIP is at an advanced stage and the IFU will
be commissioned at Calar Alto by the end of December 2003. For the spring
semester of 2004, the PPak mode of PMAS has been made available on a
shared-risk basis, requiring members of the AIP instrumentation team to be
present during the observations. For the fall semester of 2004, PPak will be
offered as a public observing mode of PMAS.

\begin{figure}
\resizebox{8cm}{8cm}
{\includegraphics[]{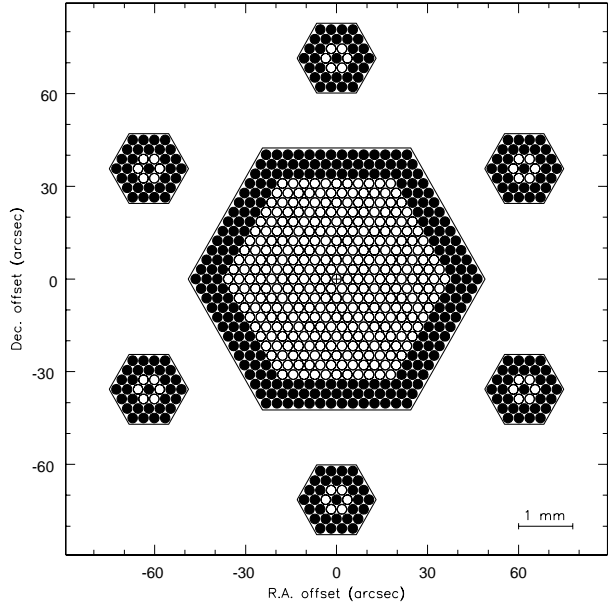}}
\caption{Layout of the 331 science and 36 sky fibers of the new fiber
  IFU for PMAS. The solid black fibers are inactive packing fibers.
}
\label{figlabel}
\end{figure}

\acknowledgements

The WIYN Observatory is a joint facility of the University of
Wisconsin-Madison, Indiana University, Yale University, and the National
Optical Astronomy Observatories. SparsePak is built with NSF grants
AST-9618849 and AST-9970780. The development of PPak is supported by the
ULTROS project under the German Verbundforschung grant 05AE2BAA/4. This work
is made possible in part by the EC under contract HPRN-CT-2002-00305 and by
NSF grant AST-0307417.


\end{document}